\documentclass[10pt, twocolumn, nofootinbib,superscriptaddress]{revtex4-1}
\usepackage{amsmath,amssymb,amsfonts}
\usepackage{algorithmic}
\usepackage{graphicx}
\usepackage{textcomp}
\usepackage{ulem}
\usepackage{xcolor}
\usepackage{ragged2e}
\usepackage{lineno}
\usepackage{booktabs, makecell, tabularx}

\usepackage{gensymb}

\makeatletter
    \renewcommand\@make@capt@title[2]{%
     \@ifx@empty\float@link{\@firstofone}{\expandafter\href\expandafter{\float@link}}%
      {\textbf{#1}}\@caption@fignum@sep#2\quad}%
\makeatother
\makeatletter 
\renewcommand{\fnum@figure}{\textbf{Fig.~\thefigure}} 
\makeatother

\usepackage{xcolor}

\def\BibTeX{{\rm B\kern-.05em{\sc i\kern-.025em b}\kern-.08em
    T\kern-.1667em\lower.7ex\hbox{E}\kern-.125emX}}

\begin{document}

\author{Chuangchuang Wei}
\thanks{These authors contributed equally to this work}
\affiliation{Nonlinear Nanophotonics Group, MESA+ Institute of Nanotechnology,\\
University of Twente, Enschede, Netherlands}
\author{Hanke Feng}
\thanks{These authors contributed equally to this work}
\affiliation{Department of Electrical Engineering and State Key Laboratory of Terahertz and Millimeter Waves, \\
City University of Hong Kong, Hong Kong, China}
\author{Kaixuan Ye}
\affiliation{Nonlinear Nanophotonics Group, MESA+ Institute of Nanotechnology,\\
University of Twente, Enschede, Netherlands}
\author{Maarten Eijkel}
\affiliation{Nonlinear Nanophotonics Group, MESA+ Institute of Nanotechnology,\\
University of Twente, Enschede, Netherlands}
\author{Yvan Klaver}
\affiliation{Nonlinear Nanophotonics Group, MESA+ Institute of Nanotechnology,\\
University of Twente, Enschede, Netherlands}
\author{Zhaoxi Chen}
\affiliation{Department of Electrical Engineering and State Key Laboratory of Terahertz and Millimeter Waves, \\
City University of Hong Kong, Hong Kong, China}
\author{Akshay Keloth}
\affiliation{Nonlinear Nanophotonics Group, MESA+ Institute of Nanotechnology,\\
University of Twente, Enschede, Netherlands}
\author{Cheng Wang}
\email{cwang257@cityu.edu.hk}
\affiliation{Department of Electrical Engineering and State Key Laboratory of Terahertz and Millimeter Waves, \\
City University of Hong Kong, Hong Kong, China}
\author{David Marpaung}
\email{david.marpaung@utwente.nl}
\affiliation{Nonlinear Nanophotonics Group, MESA+ Institute of Nanotechnology,\\
University of Twente, Enschede, Netherlands}

\date{\today}
\title{Programmable multifunctional integrated microwave photonic circuit on thin-film lithium niobate}

\begin{abstract}
Microwave photonics, with its advanced high-frequency signal processing capabilities, is expected to play a crucial role in next-generation wireless communications and radar systems. The realization of highly integrated, high-performance, and multifunctional microwave photonic links will pave the way for its widespread deployment in practical applications, which is a significant challenge. Here, leveraging thin-film lithium niobate intensity modulator and programmable cascaded microring resonators, we demonstrate for the first time a tunable microwave photonic notch filter that simultaneously achieves high level of integration along with high dynamic range, high link gain, low noise figure, and ultra-high rejection ratio. Additionally, this programmable on-chip system is multifunctional, allowing for the dual-band notch filter and the suppression of the high-power interference signal. This work demonstrates the potential applications of the thin-film lithium niobate platform in the field of high-performance integrated microwave photonic filtering and signal processing, facilitating the advancement of microwave photonic system towards practical applications.

\end{abstract}
\maketitle
\section{Introduction}
Traditional radio frequency (RF) filtering technologies are increasingly insufficient to meet the demands of future communication systems, which require large bandwidth, programmability, multifunctionality, and reductions in volume, weight, and power consumption \cite{akyildiz20206g}. Integrated microwave photonic (MWP) filters exploit the inherent broadband and flexible tuning characteristics of photonic components, combined with existing photonic integration technologies, to forge a new path for RF signal processing \cite{yao2009microwave, capmany2007microwave, marpaung2013integrated, marpaung2019integrated, capmany2006tutorial, liu2020integrated}. In recent years, significant advancements have been made in integrated MWP circuits. However, these studies often struggle to simultaneously achieve high integration \cite{fandino2017monolithic}, multifunctionality \cite{perez2017multipurpose, perez2020multipurpose, bogaerts2020programmable}, and excellent RF metrics \cite{cox2006limits, lim2009fiber, liu2018link}, which are critical for practical RF applications.

\begin{figure*}[t!]
\centering
\includegraphics[width=\linewidth]{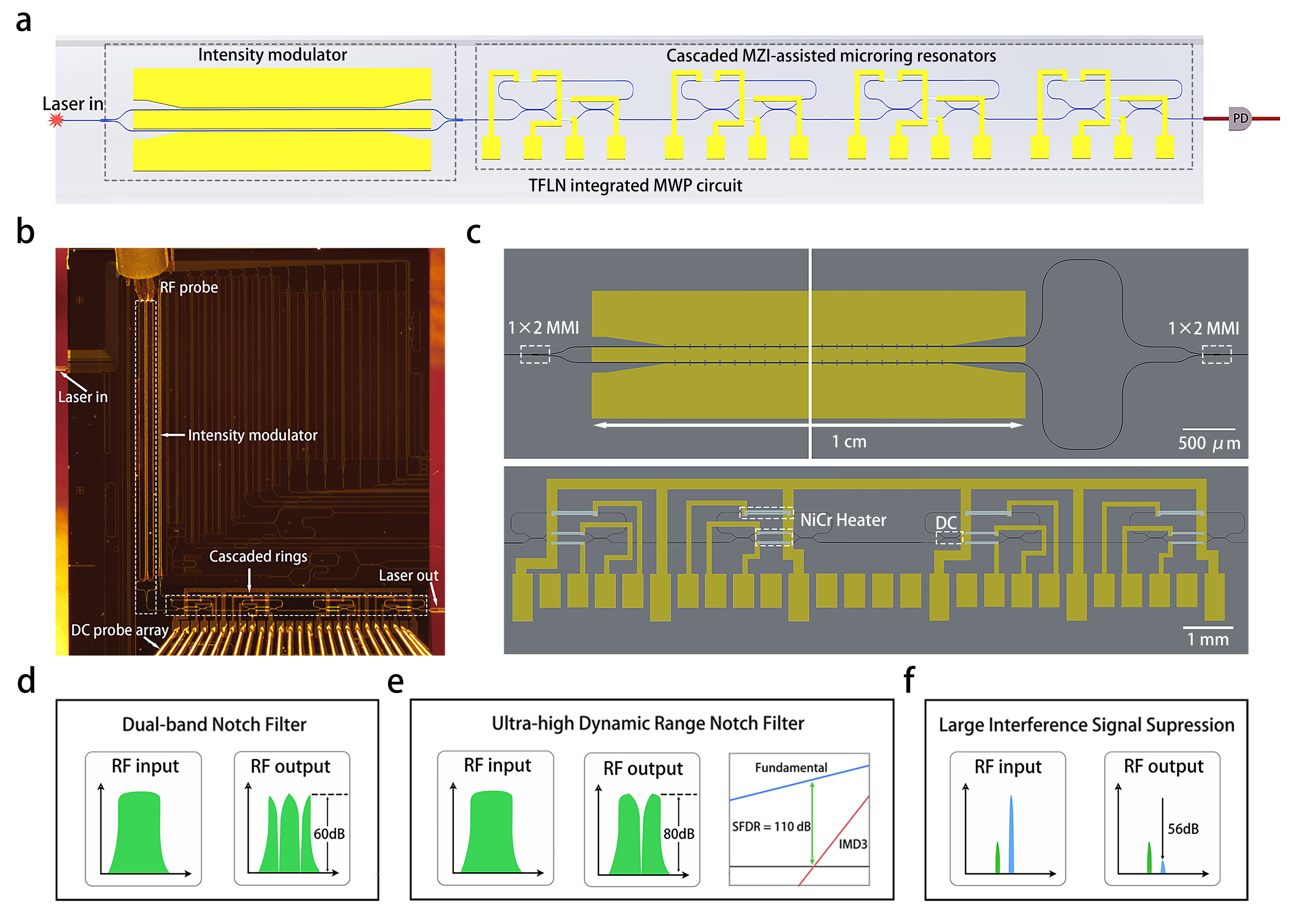}
\caption{\textbf{Multifunctional and high-Performance integrated microwave photonic circuit based on TFLN.} \textbf{a} Structure of the integrated TFLN microwave photonics chip, which consists of an intensity modulator and four cascaded Mach-Zehnder Interferometer-assisted microrings. \textbf{b} Microscope image of the chip and the experimental setup. \textbf{c} False color image of the chip. \textbf{d} Dual-band notch filter with a rejection ratio of 60 dB. e An ultra-high dynamic range notch filter with a rejection ratio of 80 dB and a SFDR of 110 dB in 1 Hz. \textbf{f} Suppression of large interference signal with a suppression ratio of 56 dB. MMI multi-mode interferometer, DC directional coupler, NiCr Nickel Chromium.}
\label{fig1}
\end{figure*}
In MWP systems, electro-optic modulators are responsible for faithful RF-to-optical conversion, and nearly all challenges stem from the modulator. The widely demonstrated MWP filters based on silicon-on-insulator (SOI) \cite{rasras2009demonstration, qiu2018continuously, tao2023highly, tao2021hybrid, zhang2022ultralow, yan2023wideband} or silicon nitride \cite{roeloffzen2013silicon, taddei2014fully, marpaung2013si, daulay2022ultrahigh, liu2017all, zhuang2015programmable} microring resonators, as well as those utilizing the stimulated Brillouin scattering (SBS) effect in waveguides \cite{casas2015tunable, marpaung2015low, gertler2022narrowband, garrett2023integrated, botter2022guided, gertler2020tunable, ye2023brillouin}, typically make use of commercial external lithium niobate modulators or compact SOI chip-based modulators for electro-optic conversion. The aforementioned approaches have limitations because commercial external lithium niobate modulators present challenges for high level of integration due to their large size, and the large half-wave voltage also limits link gain. Although silicon photonics technology has enabled the integration of modulation \cite{reed2010silicon, xu2005micrometre, li2018silicon} and MWP functions on the same photonic chip with reduced footprint, the typically higher drive voltages required by silicon modulators present challenges for linearity and power handling, which are problems particularly for MWP applications.

Thin-film lithium niobate (TFLN), with its strong and linear electro-optic effect and low waveguide propagation loss \cite{zhang2017monolithic}, represents an ideal platform for high-performance modulators \cite{wang2018integrated, mercante2018thin, wang2018nanophotonic, rao2017compact, he2019high, zhang2021integrated, xu2020high, kharel2021breaking} and other optical components \cite{zhu2021integrated, boes2023lithium, feng2024integrated}. Therefore, TFLN is an ideal platform for fabricating integrated MWP circuits, as it provides excellent electro-optic performance and supports a variety of components \cite{snigirev2023ultrafast, yu2023integrated}. However, to date, there has yet to be a demonstration of integrating TFLN modulators and programmable optical structures on the same chip to achieve multifunctional MWP circuits.

In this work, we demonstrate for the first time a highly reconfigurable TFLN MWP circuit with three distinct functionalities in a compact footprint. By integrating an intensity modulator and cascaded programmable microring resonators on a single chip, we have achieved a tunable notch filter with high rejection ratio, enhanced the spurious free dynamic range (SFDR) by suppressing third-order intermodulation distortion (IMD3), and suppressed the high-power interference signals that are very close in frequency to the signal of interest. 

The notch filter can be tuned over a frequency range of 20 GHz and can switch between single-band and dual-band operation while maintaining a rejection ratio beyond 60 dB. The linearization method suppresses IMD3 by 34 dB, increasing the SFDR of the notch filter to 110 dB in 1 Hz. Additionally, high-power interference signals that are very close in frequency to the signal of interest can be suppressed by 56 dB. Our demonstration illustrates the feasibility of achieving multifunctional and high-performance MWP circuits in a compact footprint, contributing to the application of integrated MWP technology in future communication systems.

\section{Results}

\begin{figure*}[t!]
\centering
\includegraphics[width=1\linewidth]{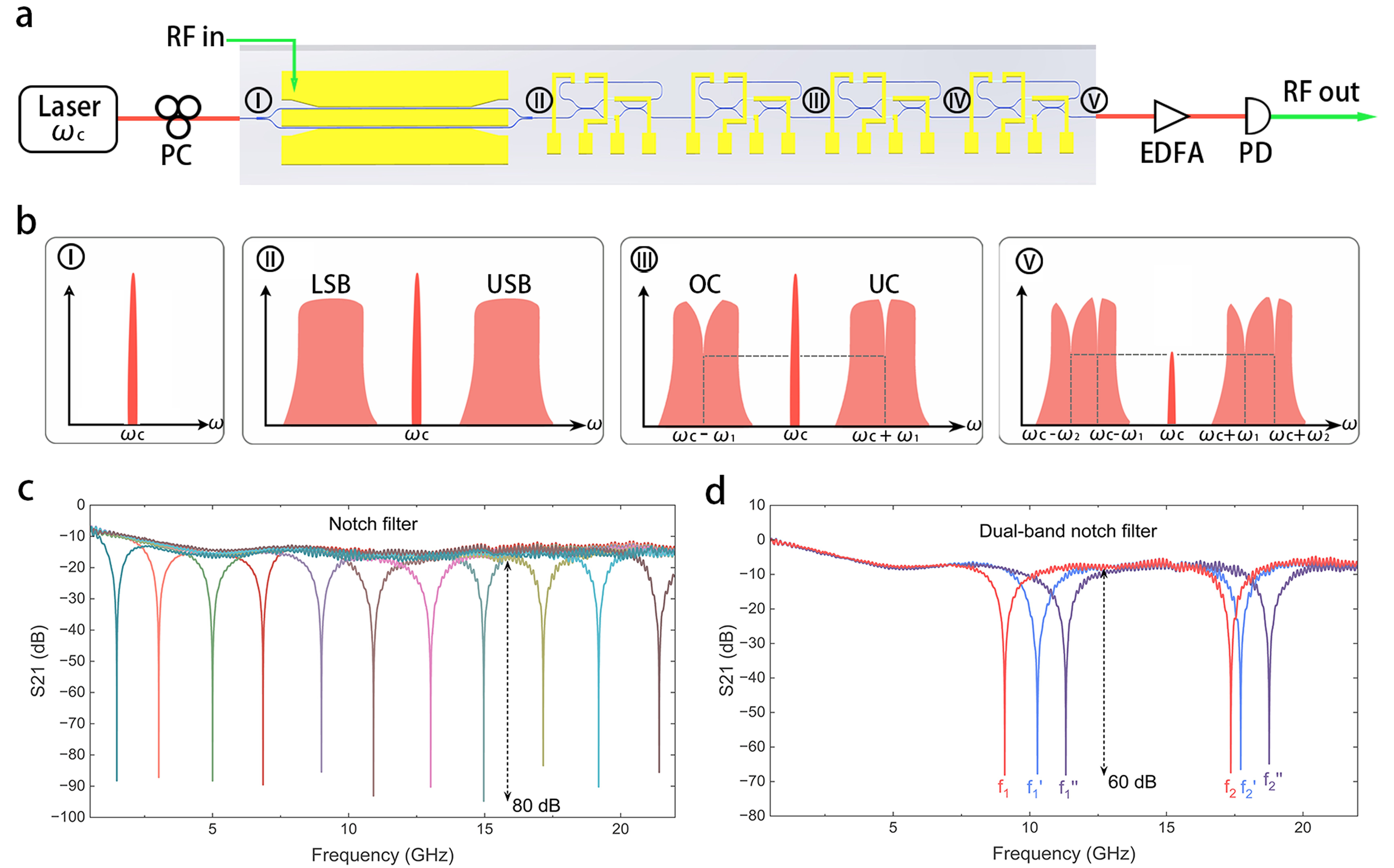}
\caption{\textbf{Operation scheme of the tunable microwave photonic notch filter.} \textbf{a} Experimental setup of programmable microwave photonic circuit, where red line represent optical path and green line represent electrical path. \textbf{b} Optical spectrum at different stage when a sweep signal is the RF input. \textbf{c} A high link gain, ultra-high rejection ratio notch filter tunable from 1.5 GHz to 21.5 GHz. \textbf{d} Tunable dual-band notch filter. LSB lower sideband, USB upper sideband, OC over-coupled, UC under-coupled.}
\label{fig2}
\end{figure*}

\begin{figure*}[t!]
\centering
\includegraphics[width=\linewidth]{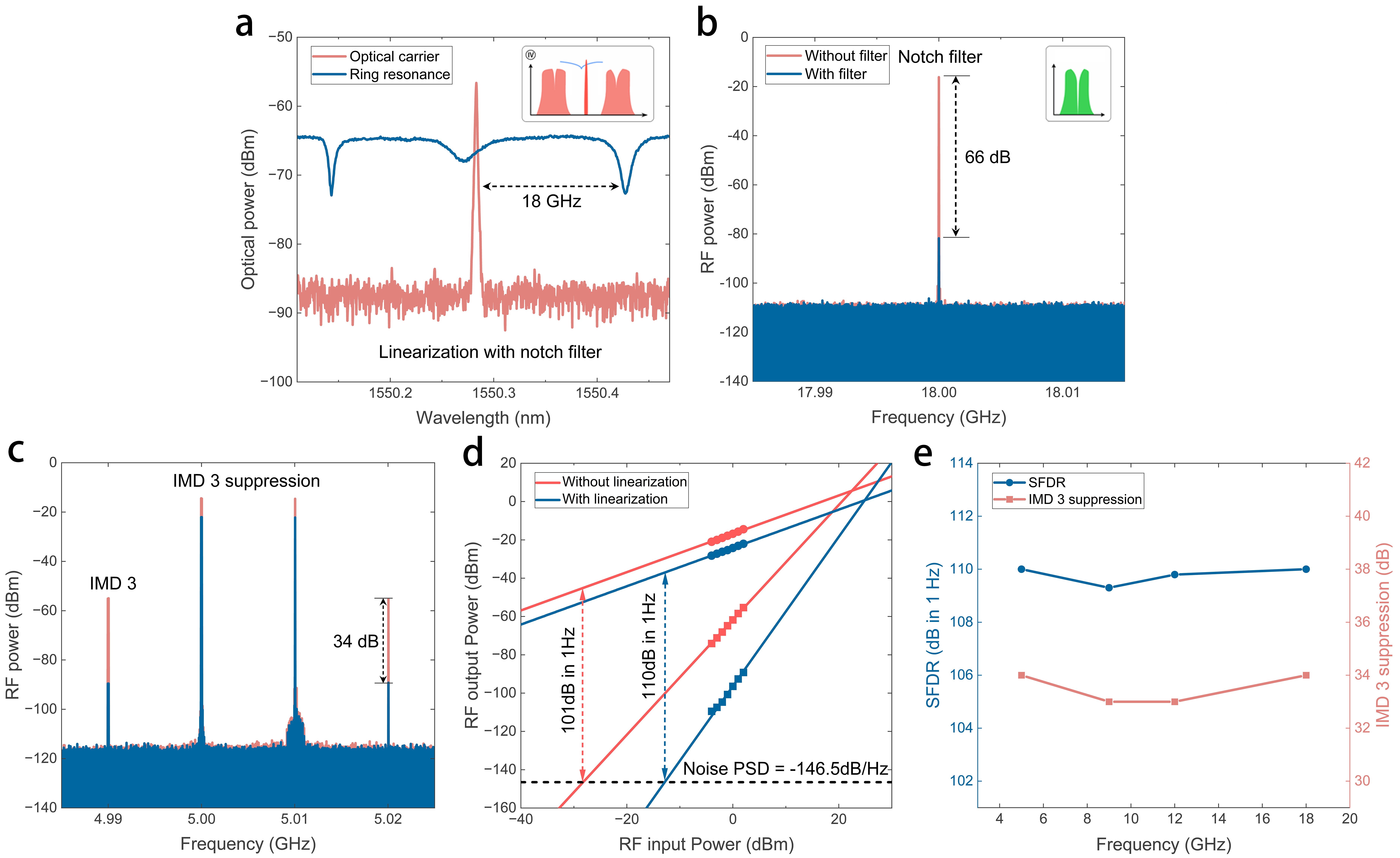}
\caption{\textbf{Ultra-high dynamic range RF notch filter.} \textbf{a} The corresponding positions of the optical carrier (red) and the microrings response (blue) when linearization is achieved. Inset: Spectrum at node 4 in Fig.2a. \textbf{b} Filtering results of the 18 GHz RF signal: the red line represents the output without filter, while the blue line represents the output with filter. \textbf{c} The RF output of the two-tone test: compared to the output without linearization (red), the output with linearization (blue) reduces IMD3 by 34 dB. \textbf{d} The measured SFDR at RF frequency of 5 GHz: the SFDR without linearization (red) is 101 dB, while with linearization (blue) it is improved to 110 dB in 1 Hz. \textbf{e} IMD3 suppression and SFDR at different frequencies.}
\label{fig3}
\end{figure*}

\begin{figure*}[t!]
\centering
\includegraphics[width=\linewidth]{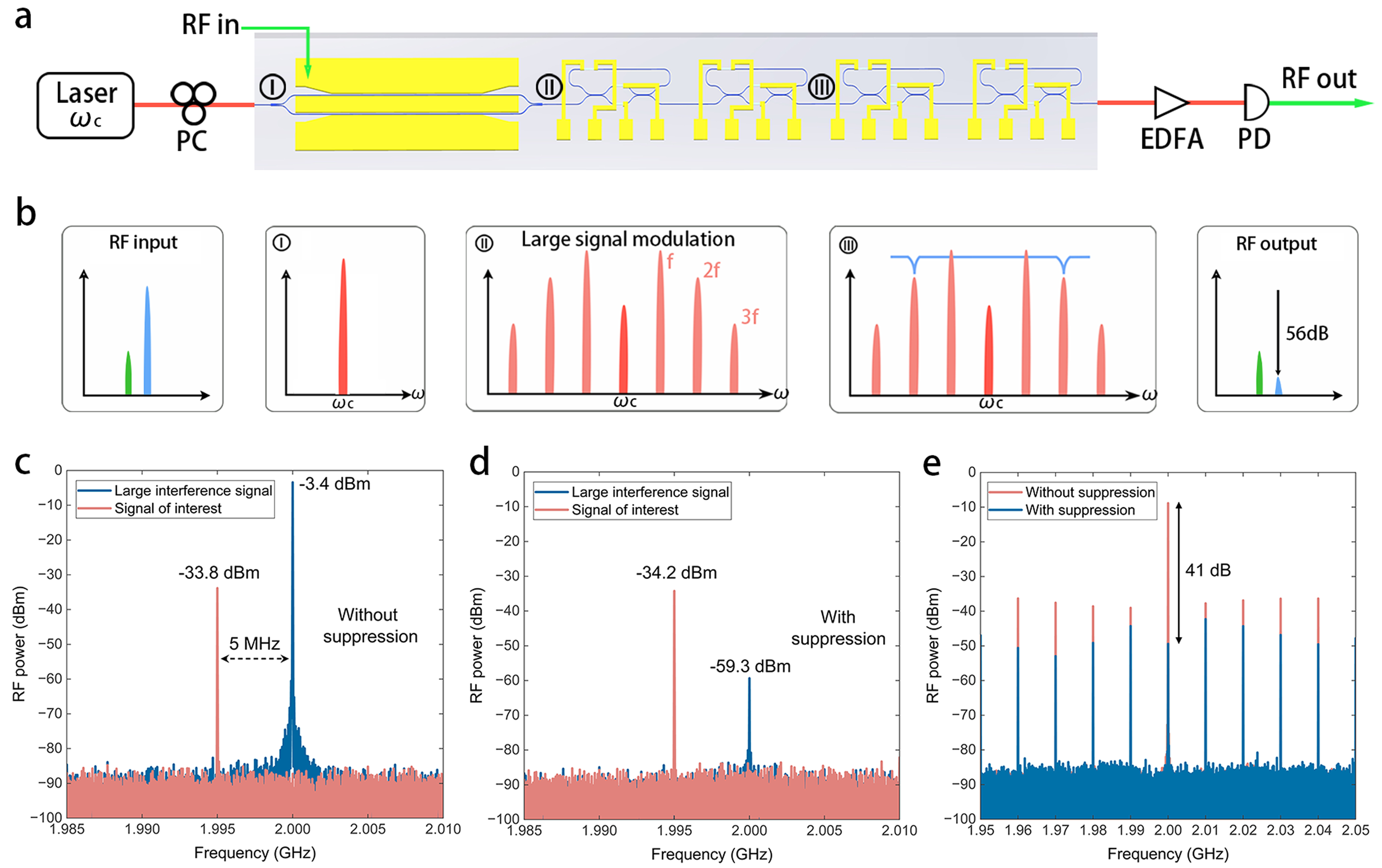}
\caption{\textbf{Large interference signal suppression.} \textbf{a} Experimental Setup of large interference signal suppression. \textbf{b} RF input and output, and optical sprctrum at different stage when achive the suppression. \textbf{c} The RF output without suppression of the large signal (blue). \textbf{d} The RF output after suppression of the large signal (blue). The signal of interest (red) remains unaffected. \textbf{e} Suppression of modulated large interference signal.}
\label{fig4}
\end{figure*}

\subsection*{Multifunctional microwave photonic circuit}

The structural diagram of the chip is shown in Fig.1a. It consists of a TFLN intensity modulator and four cascaded programmable microrings. The intensity modulator converts the RF signal into the optical domain, and the modulation sidebands are processed by Mach-Zehnder Interferometer (MZI)-assisted microrings, whose coupling coefficient and resonant wavelength can be adjusted via thermal phase shifters. This forms the basis of our multifunctional capability. Fig.1b shows a microscope image of the chip and the experimental setup. The chip employs edge coupling structure, with light coupled in and out via lensed fibers. The insertion loss is 14 dB, which can be reduced by designing a specialized taper structure \cite{he2019low}.

We load the RF signal onto the modulator electrodes using a GSG structure RF probe and control the cascaded microrings with a 26-channel DC probe array. Fig.1c is a false-color optical image of the chip. The modulator electrode is 1 cm long and adopts a slot electrode structure, with a V$\pi$ of 3.5V at 5GHz. The characterization method for RF V$\pi$ is provided in Supplementary Information \uppercase\expandafter{\romannumeral1}.
The coupling region of the microrings consists of a 3dB directional coupler and an MZI, with heaters (grey) on the MZI arms acting as phase shifters. By precisely controlling the voltage applied to the heaters on the MZI arms, each microring can be set to a specific coupling state. Heaters in the racetrack region of the microrings are used to change the resonance positions. Detailed theoretical analysis and performance characterization of the programmable cascaded microrings can be found in Supplementary Information \uppercase\expandafter{\romannumeral2}.

We have achieved three distinct functionalities based on this circuit. The first functionality is a dual-band notch filter that can simultaneously filter out any two unwanted signals with a rejection ratio of up to 60 dB, as shown in Fig.1d. The second functionality is an ultra-high dynamic range notch filter that improves the SFDR to 110 dB in 1 Hz while maintaining an 80 dB rejection ratio, as shown in Fig.1e. The third functionality can suppress high-power interference signal with frequency close to the signal of interest, as illustrated in Fig.1f.

\subsection*{Tunable notch filter}
The schematic of the entire microwave photonic link setup is shown in Fig.2a. We demonstrate ultra-high extinction MWP notch filter through phase and amplitude shaping of two modulated sidebands using two ring resonances with different coupling states \cite{liu2017all}. Key for this filter operation is to symmetrically position two ring resonances with respect to the optical carrier, where one resonance is in the overcoupling state and the other in the undercoupling state. Both resonances need to have the same extinction. Upon photodetection, the signal frequencies at the center of these resonances will create destructive RF interference leading to an ultra-high filter extinction. In experiments, we use a sweep signal generated by a vector network analyzer (VNA) as the RF input. Fig. 2b illustrates the spectra at different nodes. After the optical carrier passes through the intensity modulator, two in-phase modulation sidebands are generated, as shown at Node \uppercase\expandafter{\romannumeral2}. The spectrum after processing by the two microrings is depicted at Node \uppercase\expandafter{\romannumeral3}. It can be observed that the resonance positions of the two microrings are symmetric with respect to the optical carrier, and their suppression ratios are identical. The difference lies in the coupling states, with one microring in the undercoupled state with 0 phase shift and the other microring in the overcoupled state with $\pi$ phase shift. This meets the condition for destructive interference, leading to the ultra-high rejection ratio. Meanwhile, other parts of the modulation sidebands that are not processed by the microrings still add constructively, ensuring high link gain for the system.

Fig.2c presents the tunable notch filter response with the modulator operating at the quadrature point. The maximum link gain is -7 dB, with a suppression ratio of up to 80 dB and a 3-dB bandwidth of 1.3 GHz. This filter can be continuously tuned from 1.5 GHz to 21.5 GHz, as the tuning range is limited by the microrings' Free Spectral Range (FSR) of 44 GHz. Additionally, when the modulator is set to operate at the null point and the other two microrings are turned on, the output spectrum is
as shown at Node \uppercase\expandafter{\romannumeral5}. Based on the same working principle mentioned earlier, we achieved a tunable dual-band notch filter that can operate at two frequencies, $f_1$ and $f_2$, or can be tuned to a series of alternative frequency pairs such as $f_1$$^{\prime}$ and $f_2$$^{\prime}$, or $f_1$$^{\prime\prime}$ and $f_2$$^{\prime\prime}$, with a maximum link gain of 0 dB and a rejection ratio of up to 60 dB, as shown in Fig.2d.

In this experiment, the choice of bias point directly affects the link gain and noise figure due to the use of an intensity modulator for electro-optic conversion\cite{liu2018link}. However, since we placed the Erbium-Doped Fiber Amplifier (EDFA) directly before the photodetector (PD), the low bias technique could not optimize the noise figure, and it remained stable around 35 dB regardless of the bias voltage changes. We employed a bias tee to control the bias point, and detailed analyses of link gain and noise figure at different bias voltages are provided in Supplementary Information \uppercase\expandafter{\romannumeral3}.

\subsection*{Linearization for high dynamic range}

When two RF signals of different frequencies are input into the microwave photonic system, intermodulation distortion (IMD) occurs, limiting the system's dynamic range. We utilize the response of microrings to shape the optical carrier, thereby suppressing the third-order intermodulation distortion (IMD3) and improving the spurious-free dynamic range (SFDR). The relative positions between the optical carrier and the microring resonance peaks when achieving the ultra-high dynamic range notch filter are shown in Fig.3a. This result was obtained from an optical spectrum analyzer (OSA), which corresponds to the spectrum at Node \uppercase\expandafter{\romannumeral4} in Fig.2a. 

The responses of overcoupled and undercoupled microrings with identical suppression ratios are symmetrically distributed around the optical carrier, enabling the filtering function. The filtering result is shown in Fig.3b, where the red curve represents the RF output without filtering and the blue curve represents the filtered RF output. Concurrently, a third overcoupled microring is employed to process the optical carrier. Under specific conditions, we successfully suppressed the IMD3. (see Supplementary Information \uppercase\expandafter{\romannumeral4}) 

In the linearization demonstration, we utilized a two-tone signal generated by two signal generators as input, with the modulator operating at the quadrature point. Fig.3c illustrates the RF output from the two-tone test. It can be seen that after linearization, IMD3 is suppressed by 34 dB. From Fig.3d, it is evident that with a noise power spectral density (PSD) of -146.5 dBm/Hz in the system, the SFDR was 101 dB in 1 Hz without linearization, and it improved to 110 dB in 1 Hz after linearization. Additionally, it can be observed that after linearization, the slope of the intermodulation distortion terms has changed due to the severe suppression of IMD3, as previously insignificant fifth-order intermodulation signals start to affect the output results. To further characterize the performance of our proposed linearization scheme, we extended our tests of IMD3 suppression and SFDR at different frequencies, as shown in Fig. 3e. The results demonstrate successful linearization and IMD3 suppression at each frequency point. Detailed experimental results are provided in Supplementary Information \uppercase\expandafter{\romannumeral4}.

\begin{table*}[t!]\footnotesize
\caption{\textbf{Performance comparison of integrated MWP circuits}}
\label{tab}
\centering
\setlength{\tabcolsep}{1.03mm}{
\begin{tabular}{|lllllllllll|}
\hline
\textbf{Year} & \textbf{\begin{tabular}[c]{@{}l@{}}Material\\ platform\end{tabular}} & \textbf{\begin{tabular}[c]{@{}l@{}}Integration\\ level\end{tabular}} & \textbf{\begin{tabular}[c]{@{}l@{}}Optical\\ processing\end{tabular}} & \textbf{\begin{tabular}[c]{@{}l@{}}Function\\ number\end{tabular}}&\textbf{\begin{tabular}[c]{@{}l@{}}Tuning\\ range \\ (GHz)\end{tabular}}&\textbf{\begin{tabular}[c]{@{}l@{}}Resolution \\ (MHz)\end{tabular}}&\textbf{\begin{tabular}[c]{@{}l@{}}Performance\\ enhancement\end{tabular}}  & \textbf{\begin{tabular}[c]{@{}l@{}}Gain\\ (dB)\end{tabular}} & \textbf{\begin{tabular}[c]{@{}l@{}}Noise \\figure\\ (dB)\end{tabular}} & \textbf{\begin{tabular}[c]{@{}l@{}}SFDR\\ (dB Hz$^{2/3}$)\end{tabular}} \\ \hline

2020\cite{gertler2020tunable}      & Si & Passive & SBS & 1 & 4-10 & 3.5 & No & -17.3 & 56.7 & 93.5 
\\ \hline
2021\cite{tao2021hybrid}      & Si & MD, Passive, PD & MRR  & 2 &3-25 & 360 & No & -28.2 & 51.2 & 99.7 
\\ \hline
2022\cite{daulay2022ultrahigh}      & Si$_3$N$_4$ & Passive & DI-MRR & 6 &5-20 & 400 & SFDR, NF & 10 & 15 & 116 
\\ \hline
2022\cite{gertler2022narrowband}     & Si  & Passive & SBS   & 1 &8-14 & 2.4 & No & -3.6 & 52.5 & 93.6 
\\ \hline
2023\cite{tao2023highly}      & Si & MD, Passive, PD & MRR   & 2 &5-30 &220 & No & -27 & 47 & 92.3 
\\ \hline
2023\cite{garrett2023integrated}      & Si+As$_2$S$_3$ & MD, Passive & SBS & 2 &2.5-15 & 37 & No & -35 & 57.5 & 80        \\ \hline
This work & TFLN & MD, Passive & MRRs & 3 &1.5-21.5 & 1300/5 & SFDR & -7.2 & 34.7 & 110                                   \\ \hline
\multicolumn{11}{|l|}{\begin{tabular}[c]{@{}l@{}}MD modulator, PD photodetector, SBS stimulated Brillouin scattering, MRR microring resonator, DI double injection, \\ SFDR spurious-free dynamic, NF noise figure rang\end{tabular}}                                                                                                                                              \\ \hline

\end{tabular}}

\end{table*}
\subsection*{Large interference signal suppression}

In practical applications, the RF receiving front-end may encounter interference from high-power RF signals. When the frequency of the signal of interest and the interfering signal are very close, a filter with ultra-high resolution is needed to eliminate the interference. However, achieving extremely high-resolution MWP filters is a significant challenge. Here, we propose a method to suppress high-power interference signals. Fig.4a presents the experimental setup for demonstrating large interference signal suppression. 

Fig.4b shows the RF and optical spectrum at different nodes. When a high-power RF signal is input, high-order modulation sidebands, as shown at Node \uppercase\expandafter{\romannumeral2}, are generated. We utilize the response of the microrings to process these high-order sidebands, as illustrated by the spectrum at Node \uppercase\expandafter{\romannumeral3}. By precisely controlling the microring for shaping higher-order sidebands, the high-power interference signal can be suppressed under specific conditions (see Supplementary Information \uppercase\expandafter{\romannumeral5}). The low-power signal of interest remains unaffected because it does not generate high-order modulation sidebands. 

In this demonstration, the modulator operates at the quadrature point, with the input power of the interference signal set at 23 dBm and the input power of the signal of interest at -24 dBm. The frequency spacing between the two signals is 5 MHz, an exceptionally narrow spacing that nearly no existing MWP circuits are capable of discriminating. The RF output without suppression is shown in Fig.4c, and the RF output after suppression is shown in Fig.4d. It can be observed that only the high-power signal is suppressed, with a suppression ratio of 56 dB. Note that this large-signal suppression mechanism is based on higher-order modulation effects, that requires a minimum input RF signal strength for proper operation (see Supplementary information for the detailed theoretical analysis). For our particular TFLN modulator, the minimum RF signal strength that can be effectively suppressed was calculated to be 20.2 dBm.

To better simulate real-world radar system scenarios and demonstrate the system’s robustness, we have extended our experiments to test the suppression of modulated high-power interference signals. An arbitrary waveform generator (AWG) was used to produce multi-tone signals to simulate modulation. The generated multi-tone signal had a center frequency of 2 GHz, which was considered the carrier. Eight equidistant signals spaced at 10 MHz on both sides of the carrier were generated as modulation sidebands, with their power levels set 30 dB lower than that of the carrier. Fig.4e illustrates the RF output before and after suppression, where the red curve represents the RF output without suppression and the blue curve shows the RF output after suppression. It can be observed that high-power interference signals are suppressed by 41 dB, while lower-power signals experience less suppression.

\section{Discussion}
Table \uppercase\expandafter{\romannumeral1} summarizes the performance of state-of-the-art integrated MWP circuits. In the past three years, research on MWP circuits based on SOI platforms has successfully integrated modulators with microrings \cite{tao2021hybrid, tao2023highly} or SBS spiral lines \cite{garrett2023integrated}. However, these works have neglected improvements in RF performance, with system dynamic ranges often below 100 dB Hz$^{2/3}$. Additionally, the relatively low electro-optic conversion efficiency of SOI modulators limits system link gain and noise figure. Although our previous research on the Si$_3$N$_4$ platform achieved improvements in RF performance and multifunctionality \cite{daulay2022ultrahigh}, the demonstration still relied on off-chip modulators, and the integration level remains to be enhanced. In contrast, our current work demonstrates high degree of integration, advanced functionality, and high RF performance metrics for the first time via a novel TFLN platform that consists of high speed modulator and a network of ring resonators. To date, no investigation of enhancement of RF performance metrics of MWP subsystem ever reported in TFLN. Naturally, our work shed light into the limit of performance that can be achieved in such novel systems.

Several techniques can be applied to further enhance the performance of this system. Firstly, our current structure comprises only four cascaded microrings, limiting our ability to shape the optical carrier for linearization while implementing the dual-band notch filter. In future designs, we plan to increase the number of programmable microrings, enabling the simultaneous realization of multiple functionalities. Secondly, the edge-coupling loss in the waveguide is relatively high, impacting the system's noise figure. In future work, we can address this by designing tapers to reduce the coupling loss between the fiber and the TFLN waveguide. Thirdly, we currently rely on an external EDFA to achieve high link gain, and its placement directly before the photodetector (PD) prevents us from applying the low-bias method to reduce the noise figure. A potential solution in future research could be the development of a TFLN waveguide amplifier \cite{liu2022photonic, bao2024erbium}. Additionally, in future work, we plan to leverage the electro-optic effect to more precisely control the microring response, enabling further improvements in linearization performance. 

In summary, we have designed and fabricated a programmable MWP circuit based on TFLN intensity modulator and cascaded microrings. Our experimental demonstrations highlight its multifunctionality, including high link gain, ultra-high dynamic range, ultra-high rejection ratio notch filter, as well as dual-band notch filter and suppression of large interference signals. This work marks the first demonstration of a multifunctional MWP circuit on the TFLN platform, and the improvement strategies outlined above provide a clear pathway toward achieving even higher performance. These advancements represent a significant breakthrough for the application of MWP technology in real-world communication and radar systems.

\section{Methods}
\subsection*{Device fabrication}

The MWP circuit is fabricated from commercially available x-cut lithium niobate on insulator (LNOI) wafer (NANOLN), consisting of a 500 nm thick LN thin film, a 4.7 $\mu$m thick buried SiO$_2$ layer, and a 500 $\mu$m thick silicon substrate. Firstly, SiO$_2$ is deposited on the surface of a 4-inch LNOI wafer as an etching hard mask using plasma-enhanced chemical vapor deposition (PECVD). Various functional devices are patterned on the entire wafer using an ASML UV Stepper lithography system die-by-die with a resolution of 500 nm. Next, the exposed resist patterns are transferred first to the SiO$_2$ layer using a standard fluorine-based dry etching process, and then to the LN device layer using an optimized Ar$^+$ based inductively coupled plasma (ICP) reactive-ion etching process. The etching depth of the LN layer is 250 nm, leaving a 250 nm-thick slab. After the removal of the residual SiO$_2$ mask and redeposition, an annealing process is carried out. Following the deposition of the SiO$_2$ cladding using PECVD, the microwave electrodes, heater, and wires/pads are fabricated through the second, third, and fourth lithography and lift-off processes, respectively.

\subsection*{Details of experiments}

We use a laser (Teraxion) with relative-intensity noise below -160 dBc/Hz as the optical carrier. After passing through a polarization controller, polarization beam splitter, and polarization rotator, 16.8 dBm of TE mode light is coupled into the TFLN chip. In the tunable filtering experiment, a sweep signal generated by VNA (Keysight 5007A) drives the intensity modulator. The optical signal output from the chip is amplified by a low-noise EDFA (Amonics) and then sent to a PD (APIC 40GHz) and the converted RF signal is measured with a VNA to obtain the S$_2$$_1$ curve.

In the linearization experiment, we use a two-tone RF signal with a center frequency of 5 GHz and a space of 10 MHz as the RF input, generated by two signal generators (Wiltron 69147A and HP 8672A). The RF signal output from the PD is analyzed using an RFSA (Keysight N9000B).

In the large interference signal suppression experiment, the RF signal generated by a signal generator is amplified by an RF amplifier (ZVA-213-S+) to drive the intensity modulator. The RF output is measured with an RFSA (Keysight N9000B). 

All the aforementioned experiments require the chip to operate in specific states. We use a multi-channel power supply with a precision of 0.001V to accurately apply voltage to each heater. Additionally, a thermo electric cooler (TEC) is employed to control the temperature, ensuring more stable operation.

\section*{acknowledgments}
The authors acknowledge funding from the European Research Council Consolidator Grant (101043229 TRIFFIC), Nederlandse Organisatie voor Wetenschappelijk Onderzoek (NWO) Start Up (740.018.021), Photon Delta National Growth Fund programme, the Research Grants Council, University Grants Committee (N\_CityU11320, CityU11204022), and Croucher Foundation (9509005).

\section*{Author Contribution}
Cc.W. and H.F. contributed equally to this work. Cc.W., K.Y. and D.M. proposed the concept and designed the experimental plan. H.F., K.Y., D.M. and C.W. designed and fabricated the TFLN photonic circuit. Cc.W. performed the experiments with input from D.M., K.Y., Y.K., A.K. and M.E.. Cc.W., D.M., H.F. and C.W. wrote the manuscript with input from all authors. D.M. supervised the project.

\section*{Competing interests}
The authors declare no conflicts of interest.

\section*{Data Availability}
The data that support the plots within this paper and the supplementary materials are available at https://doi.org/10.5281/zenodo.14065933.
\nolinenumbers
\bibliographystyle{IEEEtran}
\bibliography{reference}

\end{document}